\documentclass{article}

\usepackage{arxiv}

\usepackage[utf8]{inputenc} 
\usepackage[T1]{fontenc}    
\usepackage{hyperref}       
\usepackage{url}            
\usepackage{booktabs}       
\usepackage{amsfonts}       
\usepackage{nicefrac}       
\usepackage{microtype}      
\usepackage{lipsum}
\usepackage{graphicx}
\usepackage{float}
\usepackage{subfig}

\title{Segmentation of brain tumor on magnetic resonance imaging using a convolutional architecture}

\author{
 Miriam Zulema Jacobo Gomez \\
  School of Biomedical Engineering\\
  Universidad Autónoma de Ciudad Juárez\\
  Cd Juárez, Mex. 32310. \\
  \texttt{al131384@alumnos.uacj.mx} \\
   \And
 Jose Mejia \\
  School of Biomedical Engineering\\
  Universidad Autónoma de Ciudad Juárez\\
  Cd Juárez, Mex. 32310. \\
  \texttt{jose.mejia@uacj.mx} \\
}

\begin{document}
\maketitle
\begin{abstract}
The brain is a complex organ controlling cognitive process and physical functions. Tumors in the brain are accelerated cell growths affecting the normal function and processes in the brain. MRI scans provides detailed images of the body being one of the most common tests to diagnose brain tumors. The process of segmentation of brain tumors from magnetic resonance imaging can provide a valuable guide for diagnosis, treatment planning and prediction of results. Here we consider the problem brain tumor segmentation using a Deep learning architecture for use in tumor segmentation. Although the proposed architecture is simple and computationally easy to train, it is capable of reaching $IoU$ levels of 0.95.
\end{abstract}


\section{Introduction}
Brain tumors are caused by an uncontrolled growth and dissemination of cells and are classified as benign or low grade (grade I and II) and malignant or high grade (grade III and IV) \cite{Mohsen}. Although surgery is the most common treatment for brain tumors, radiation and chemotherapy can be used to slow the growth of tumors that cannot be physically removed.
Magnetic resonance imaging (MRI) provides detailed images of the brain and is one of the most common tests used to diagnose brain tumors \cite{Havaei}. Magnetic resonance imaging has a great impact in the field of medical imaging analysis because of its ability to provide much information about brain structure and abnormalities in brain tissues due to the high resolution of the images \cite{Mohsen}. Usually, an expert radiologist uses the MRI technique to generate a sequence of images to identify different regions of the tumor. This variety of images helps radiologists and experts extract different types of information about the tumor (shape, volume, etc.). Manual segmentation in MRI is a time-consuming procedure and depends on the skills of each expert \cite{Ben}. In the specific case of brain tumor, segmentation consists of the separation of different tumor tissues as an active site, edema and necrosis of normal brain tissues (gray matter, white matter and cerebrospinal fluid). In the semi-automatic segmentation of brain tumors, the intervention of a human operator is needed to initiate the method, verify the accuracy of the result or even to manually correct the segmentation \cite{Gordillo}.
In fully automatic brain tumor segmentation methods, user interaction is not required. Mainly, artificial intelligence and prior knowledge combine to solve the segmentation problem \cite{Isin}. Image processing techniques are used to extract features that represent each different type of tissue \cite{Isin}. Convolutional neural networks have the advantage of automatically learning complex characteristics representative of both healthy brain tissues and tumor tissues directly from magnetic resonance imaging. Although there are various techniques for brain tumor segmentation in the literature, deep learning techniques have proven accurate for use in tumor segmentation \cite{Mohsen,Ben}, , and accuracy can be improved by designing an architecture in which you work with deep learning and convolutional neural networks.

There are several works that study brain tumor segmentation, for example in \cite{Havaei} an automatic method of segmentation of brain tumors based on deep neural networks, two types of architectures are described: a two-way architecture, a path with $7\times 7$ receptive fields and another with larger input fields, of $13\times 13$, obtaining characteristics of both ways feeding the output layer; waterfall architecture, in which three architectures linking the exit of the first CNN to different levels of the second CNN were investigated, the two-way architecture was first trained with the two-phase stochastic gradient descent procedure described in the work, and then the training was applied to the waterfalls. As for preprocessing, a simple method based on connected components was implemented to eliminate flat spots in the predictions. With these implementations, segmentation was obtained in 25 s, so it was 45 times faster than extracting a patch in each pixel and processing it individually for the entire brain.

In \cite{Zhang} a method of encoding cerebral symmetry in a neural network is presented for a brain tumor segmentation task, they  compared to the standard U-Net network for the brain tumor segmentation task. The method consists of coding the symmetry through two ways of entry, on the one hand, the image of the normal brain enters and on the other the inverted image of the brain, the difference between the characteristic maps is calculated after several layers. This method forces the network to pay more attention to the high-level asymmetric portions, which are more likely to be tumor regions. 
In \cite{Mohsen} it is proposed an efficient methodology that combines the discrete wavelet transform with deep neural networks to classify brain magnetic resonances in normal and 3 types of malignant brain tumors: glioblastoma, sarcoma and metastatic bronchogenic carcinoma, the images were taken from the database of the Harvard University website. It uses fuzzy C-means to perform the segmentation of the tumors and in the end the network classifies what type of tumor it is and obtains an accuracy of 97\%. 

In \cite{Ben} three fully automatic methods (2CNet, 3CNet, EnsembleNet) are proposed for the segmentation of brain tumors using CNN, in which the EnsembleNet model is used to fusion the models mentioned above. The proposed deep learning work without any post-processing operations and achieve in average 0.88 Dice score.

In this work we proposed a new architecture to segment tumors in the brain, for this end we used a three channel deep convolutional network. Each channel has a receptive field of different size,  in order to extract features at different resolution. Each channel has reduced complexity, the first channel consisting of just one convolutional layer. Although the proposed architecture is simple and computationally easy to train, it is capable of reaching $IoU$ levels of 0.95.

\section{Methods}

\begin{figure}
    \centering
    \includegraphics[scale=0.7]{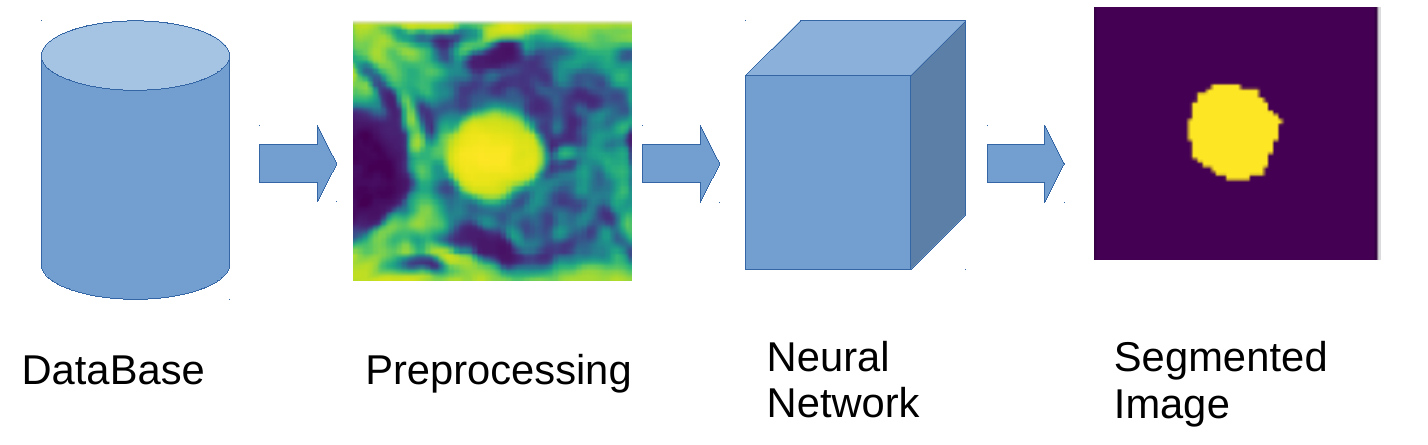}
    \caption{block diagram of the proposed architecture.}
    \label{fig:bloques}
\end{figure}

In this section the steps of the methodology are described: the preprocessing of the image, the design of the network architecture, the training, and validation of the network. In Figure \ref{fig:bloques} is shown the complete system.

{}

\subsection{Database}
The database used was obtained from the Figshare site \cite{Cheng}. It contains a total of 3064 MRI contrasted T1-weighted images of 233 patients. The tumors are divided into three different types arranged in sections: 1426 glioma sections, 708 meningioma sections and 930 pituitary tumor sections. The images were provided by Jun Cheng, a teacher and researcher at the biomedical engineering school with a focus on medical image processing at Southern Medical University, Guangzhou China.
The images of the data base are in  Matlab\textsuperscript{\textcopyright} format (.mat), and have a size of $512\times 512$.

\subsection{Image pre-processing}
The database consisted of MRI images and ground truth binary masks for the tumor. Since image size reduction is a widely used practice to save processor resources, and has been used on some works about segmentation of brain tumors \cite{Zhang}, our first step the was to reduce the size of the RMI images, this was done by only selecting the region of interest, the tumor area was delimited using a window of size $100\times 100$. This window is the one that will finally be used as an input to the network.
For the output we resize the binary mask to a size of $100\times 100$, so that the training of the network was fast and not spend on memory resources.

A database package containing a total of 1440 images was used, of which $80$\% were used as a training set and $20$\% as a test set.

\subsection{Network Architecture}

\begin{figure}
    \centering
    \includegraphics[scale=1.1]{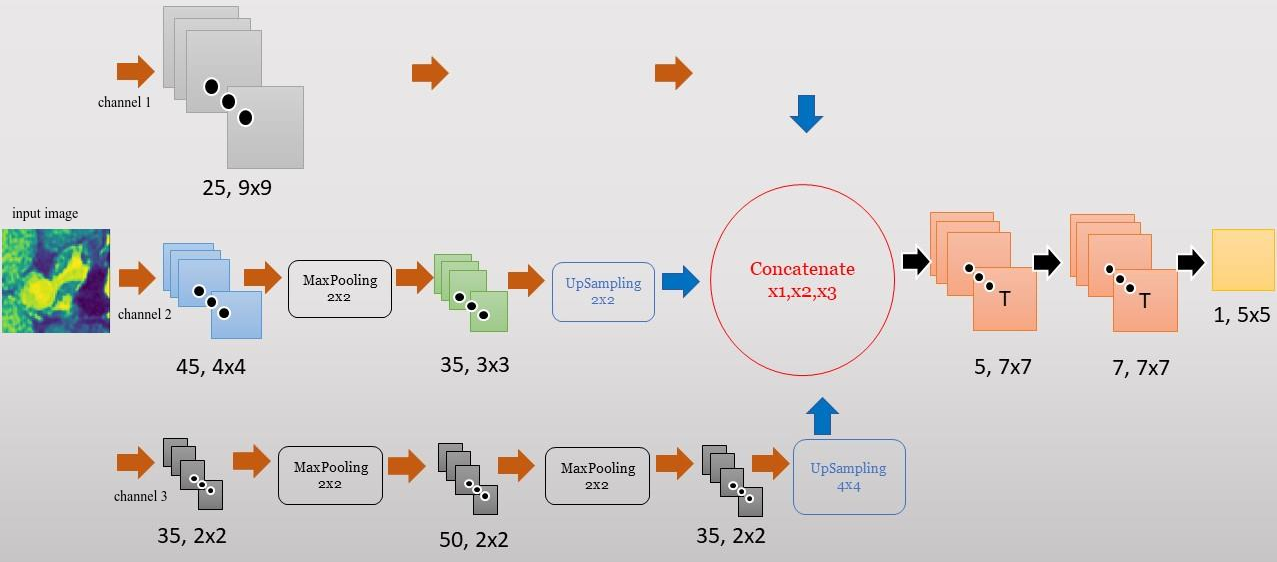}
    \caption{The proposed architecture.}
    \label{fig:red}
\end{figure}

The proposed architecture is a modification of the network used in \cite{victor}, Figure \ref{fig:red}, shows the network architecture. The designed network has three channels, through which the image to be segmented enters:
\begin{itemize}
    \item Channel 1, consists of  $25$ convolutional filters, each filter of size $9\times 9$ and this goes to a concatenate layer;
    \item Channel 2 has two convolution layers, the first is a convolutional layer of  $45$ filters of size $ 4 \times 4 $, then its output goes through a $2\times 2$ maxpooling layer, and then to the second convolution layer of $35$,  $3\times 3$ size filters following by an up sampling of size $2\times 2$, which increases the size of the output by filling with zeros before it goes to a cancatenate layer.
    \item Channel 3 has three coonvolutional layers, the first consists of 35 filters of size $2\times 2$, followed by a $2\times 2$ maxpooling, the next layer has $50$ filters of size $2\times 2$, also also followed by a maxpooling, and the last layer in the channel has 35 filters of 2x2, entering to a $4\times 4$ up sampling function before moving on to the concatenate function.
\end{itemize}{}
It is expected that each channel, having a receptive field of different size,  extract features at different resolution that helps in the segmentation process. 
All three channels are feed to a concatenate function.
After the concatenate function, two layers of transposed convolution follow, the first layer with $5$ filters of size $7\times7$, the second layer consist of $7$ filters with size $7\times7$. Finally, the last layer of the architecture is a convolution layer of one filter with size of $5\times5$ that will output the segmented MRI image.
{}

\subsection{Metrics}\label{MetricasOK}

We used the intersection on the Union (IoU) as the evaluation metric. Given an object, the IoU metric offers a measure of similarity between the predicted region of the object and the  ground-truth  region, IoU  is  defined  as the  number of pixels in the  intersection  divided  by  the number of pixels in the  union  of  the  two  regions \cite{Berman,Rahman}.\\
Additionally, we  calculate the True Positive pixels (TP) as the pixels predicted correctly as being part of the tumor, False Positive pixels (FP), as those pixels incorrectly predicted as as being part of the tumor, and False Negative pixels (FN) as pixels incorrectly classified as part of normal tissue.
We also evaluate the true positive rate (TPR) and the positive predictive value (PPV) \cite{Powers}.

\section{Results}
\begin{figure}
    \centering
    \includegraphics[scale=1.1]{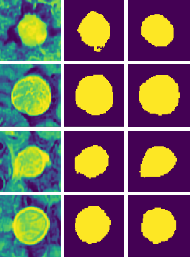}
    \caption{Output of the proposed architecture. First column original image, second column output of the proposed architecture, third column Ground truth.}
    \label{fig:resultados1}
\end{figure}

In this Section it is presented the results obtained during the execution of the brain tumor segmentation of over the test set, Figure \ref{fig:resultados1} show those results for several images of the set.

{}

\begin{figure}
    \centering
    \includegraphics[scale=0.99]{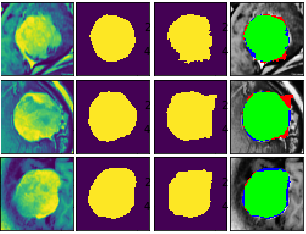}
    \caption{Results of the metrics. First column original image, second column Ground truth, third column output of the proposed architecture, and fourth column metric results: green TP, red FP, and blue FN.}
    \label{fig:metricas}
\end{figure}

\begin{figure}
 \centering
  \subfloat[]{
   \label{f:redimen}
    \includegraphics[width=0.37\textwidth]{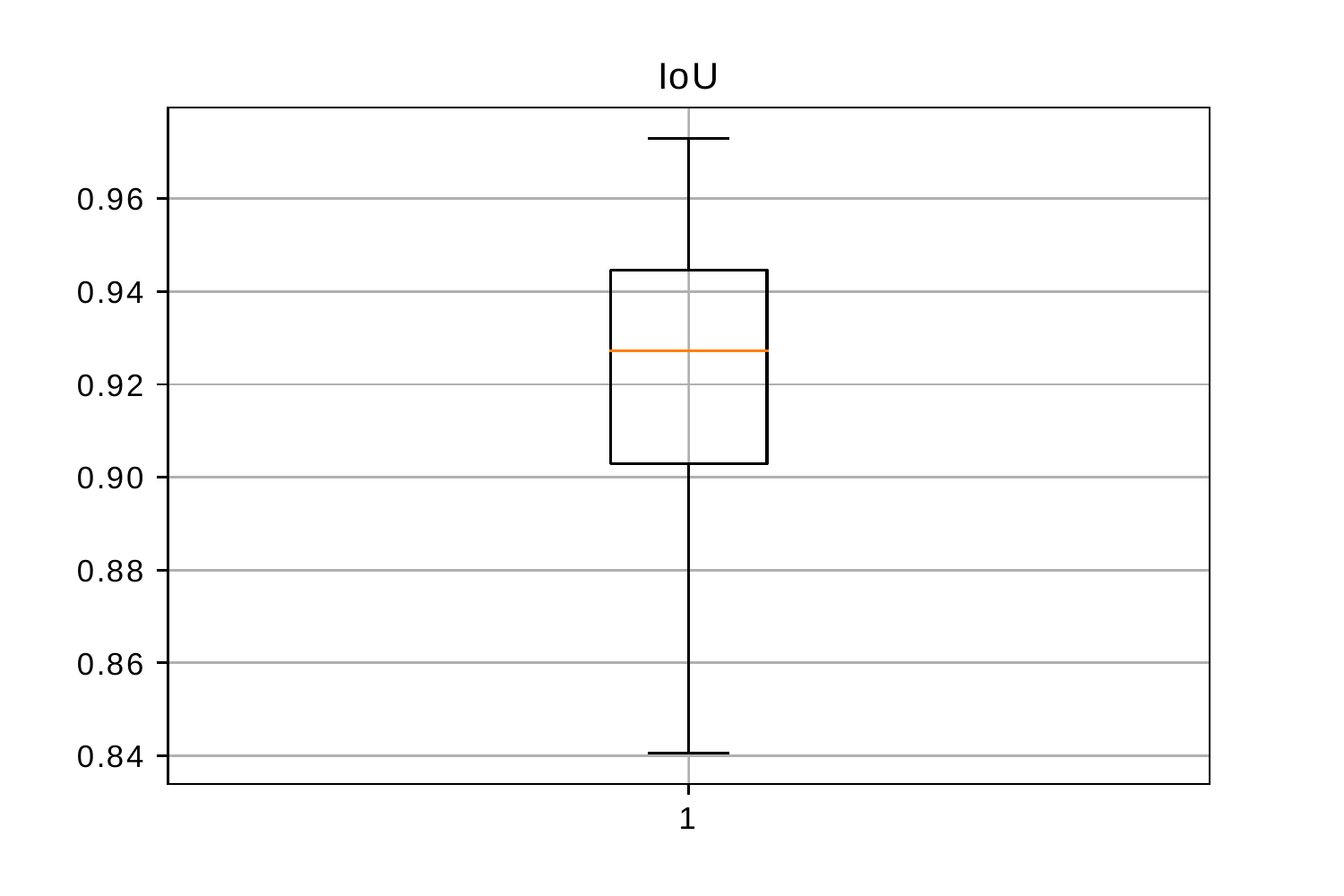}}
   \qquad 
  \subfloat[]{
   \label{f:zona}
    \includegraphics[width=0.37\textwidth]{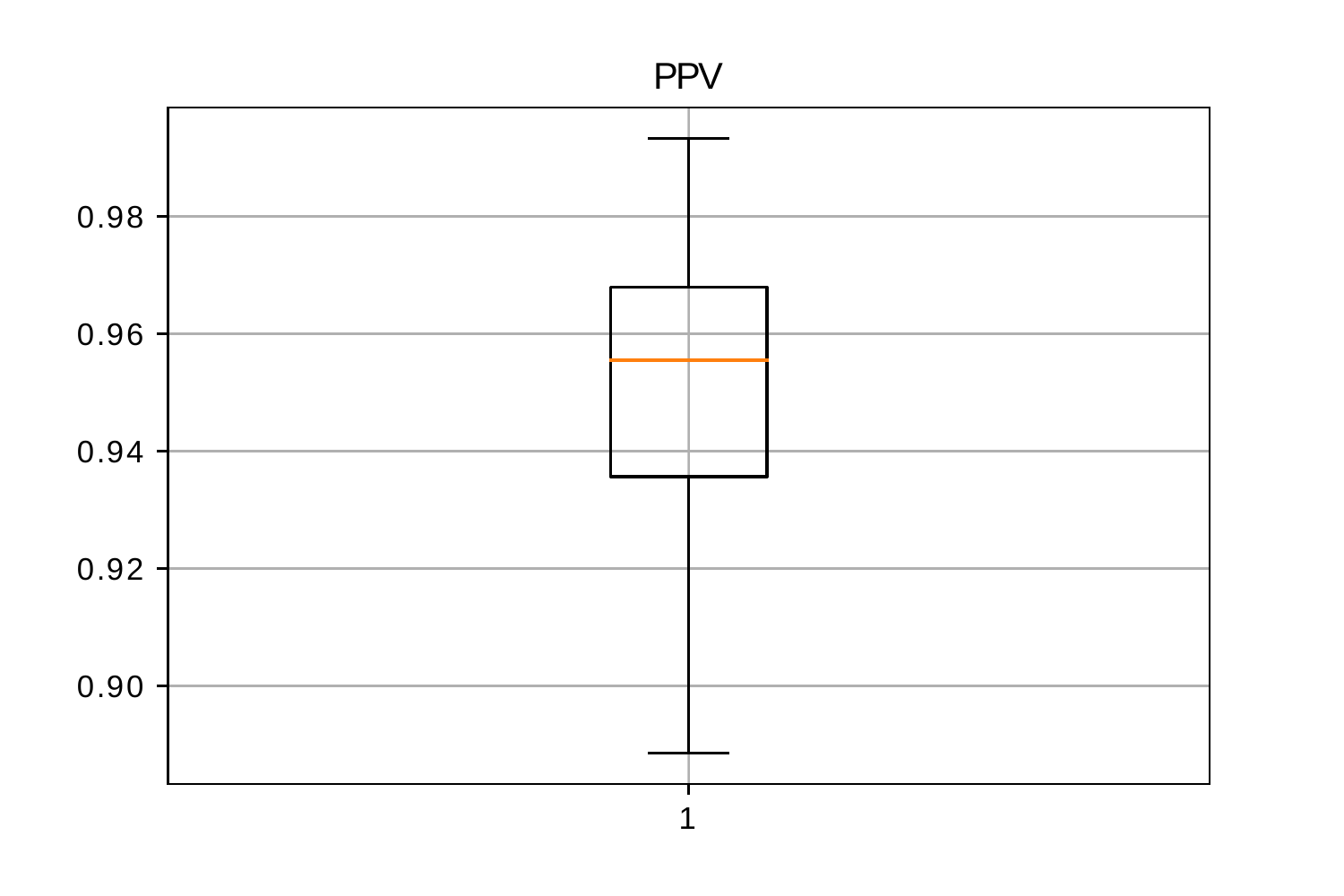}}
    \qquad
   \subfloat[]{
   \label{f:zona}
    \includegraphics[width=0.37\textwidth]{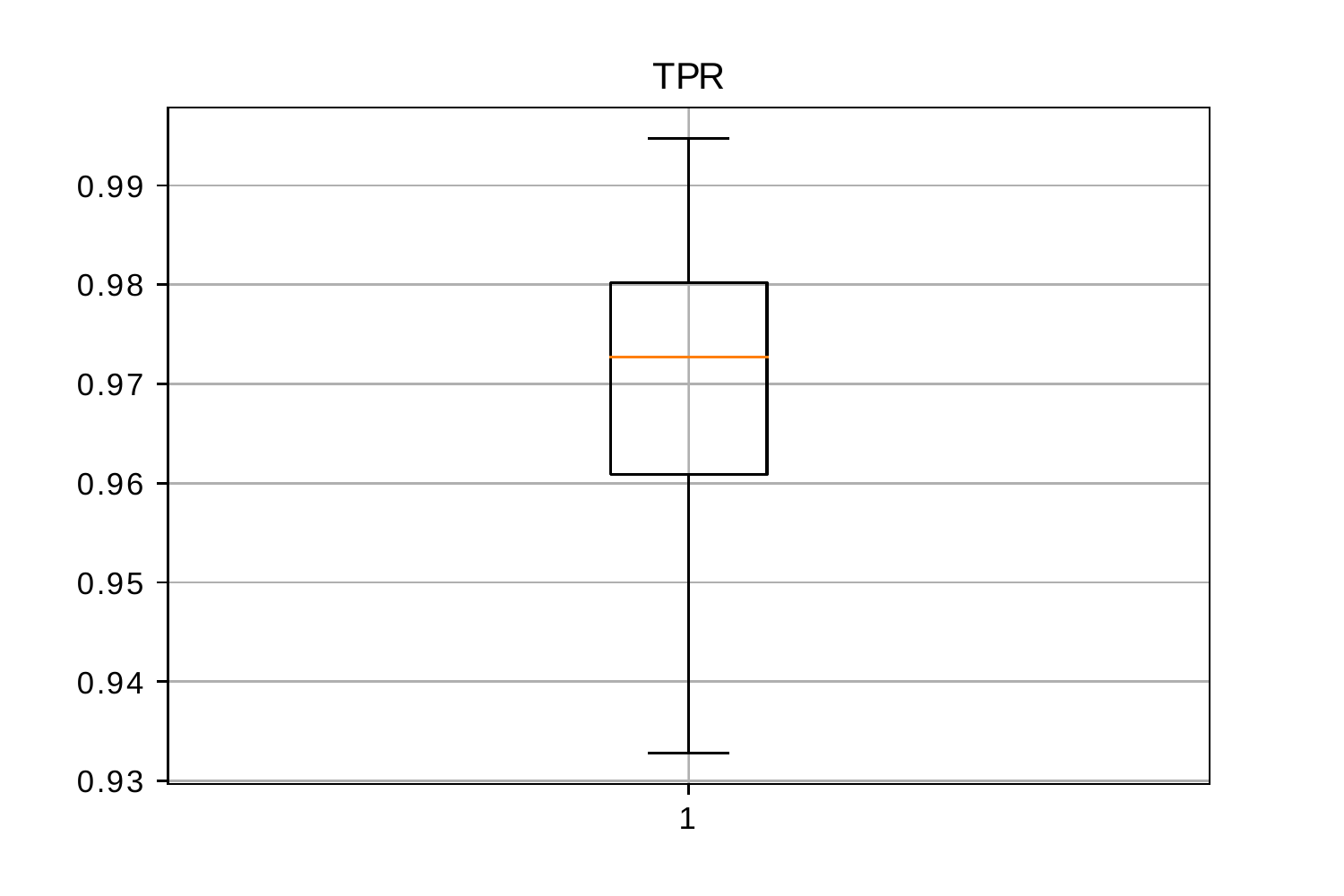}}     
    
 \caption{Boxplots of the metrics, (a) IoU, (b) PPV, and (c) TPR.}
 \label{fig:SeriesBS}
\end{figure}

In Figure \ref{fig:metricas}, it is presented, the evaluation of the metrics to the output of the architecture, for the image in the first row it is obtained an IoU=0.89, TP= 1053,
FP= 79, and FN= 57; for the second row IoU=0.87, TP= 1220, FP= 124, and FN= 56; and for the third row an IoU= 0.89, TP= 1181, FP= 14, and FN= 136.\\
Also, from \ref{fig:SeriesBS} it can be seen that the proposed architecture is capable of identify the tumor and correctly label most of its area, which is show as green, pixels incorrectly classified are the less, this is show quantitatively in the graphs of Figure \ref{fig:SeriesBS}, which presents a global summary of the metrics over all the test set. From Figure \ref{fig:SeriesBS}a, the median of the IoU distribution is roughly $0.93$  and about $50$ percent of the predictions fall inside an IoU value of $0.905$-$0.945$. Similarly, Figure \ref{fig:SeriesBS}b,c show a median of $0.955$ and $0.975$  for the PPV and TPR respectively. \\

{}

\section{Conclusions}
In this work, the segmentation of meningioma-like brain tumor, glioma and pituitary tumor from a database of 233 patients was presented, the images were preprocessed to reduce their size to 100 pixels. The architecture of the network has three channels through which the MRIs enter,  each channel are joined in a concatenation whose output is feed to another 3 layers with smaller from which the output is taken.

A validation of the segmentation  was performed using IoU to rate how much of the segmentation area is actually tumor. Our proposed architecture  is capable of reaching $IoU$ levels of 0.95.

The segmentation of brain tumors through deep learning that was developed in this paper, was carried out semi-automatically since the region of interest is manually selected.//

As future work, we planned to test the architecture with other databases or even other more complicated types of tumors.//

\bibliographystyle{unsrt}  


\begin{thebibliography}{1}

\bibitem{Mohsen}
Mohsen, Heba, El-Sayed A. El-Dahshan, El-Sayed M. El-Horbaty, and Abdel-Badeeh M. Salem.
\newblock Classification using deep learning neural networks for brain tumors.
\newblock In {\em Future Computing and Informatics Journal}, pages 68--71. 2018.

\bibitem{Havaei}
M. Havaei et al., “Brain tumor segmentation with Deep Neural Networks,” Med. Image Anal., vol. 35, pp. 18–31, 2017.

\bibitem{Ben}
M. Ben naceur, R. Saouli, M. Akil, and R. Kachouri, “Fully Automatic Brain Tumor Segmentation using End-To-End Incremental Deep Neural Networks in MRI images,” Comput. Methods Programs Biomed., vol. 166, pp. 39–49, 2018.

\bibitem{Gordillo}
N. Gordillo, E. Montseny, and P. Sobrevilla, “State of the art survey on MRI brain tumor segmentation,” Magn. Reson. Imaging, vol. 31, no. 8, pp. 1426–1438, 2013.

\bibitem{Isin}
A. Isin, C. Direkoglu, and M. Sah, “Review of MRI-based Brain Tumor Image Segmentation Using Deep Learning Methods,” Procedia Comput. Sci., vol. 102, no. August, pp. 317–324, 2016.
\bibitem{Zhang}
H. Zhang, X. Zhu, and T. L. Willke, “Segmenting Brain Tumors with Symmetry,” no. November, 2017.
\bibitem{Cheng}
Cheng, Jun (2017): brain tumor dataset. figshare. Dataset. https://doi.org/10.6084/m9.figshare.1512427.v5.

\bibitem{victor}
Victor Mendoza Guzmán, Jose Mejia, Nayeli Moreno Márquez, Paula Rodrıguez, and Santiago Ramırez. "Disparity map estimation with deep learning in stereo vision.",Regional Consortium for Foundations, Research and Spread of Emerging Technologies in Computing Sciences, 2018.

\bibitem{Rahman}
Rahman, Md Atiqur, and Yang Wang. "Optimizing intersection-over-union in deep neural networks for image segmentation." In International symposium on visual computing, pp. 234-244. Springer, Cham, 2016.
\bibitem{Berman}
Berman, Maxim, Amal Rannen Triki, and Matthew B. Blaschko. "The lovász-softmax loss: a tractable surrogate for the optimization of the intersection-over-union measure in neural networks." In Proceedings of the IEEE Conference on Computer Vision and Pattern Recognition, pp. 4413-4421. 2018.

\bibitem{Powers}
Powers, David Martin. "Evaluation: from precision, recall and F-measure to ROC, informedness, markedness and correlation." (2011).

\end{thebibliography}

\end{document}